\documentstyle{mn}
\input epsf

\def\spose#1{\hbox to 0pt{#1\hss}}
\newcommand{\approxlt}{\mathrel{\spose{\lower 3pt\hbox{$\sim$}}
	\raise 2.0pt\hbox{$<$}}}
\newcommand{\approxgt}{\mathrel{\spose{\lower 3pt\hbox{$\sim$}}
	\raise 2.0pt\hbox{$>$}}}

\newcommand {\droro}         {\delta\rho/\rho}
\newcommand {\OmegaB}        {\Omega_{\rm B}}
\newcommand {\Omegaeff}      {\Omega_{\rm eff}}
\newcommand {\Omegaoeff}     {\Omega^{\rm S}_{\rm eff}}
\newcommand {\deltaiclus}    {\Delta I_{\rm clus}}
\newcommand {\deltaioclus}   {\Delta I_{\rm Sclus}}
\newcommand {\deltainoise}   {\Delta I_{\rm noise}}
\newcommand {\deltaionoise}  {\Delta I_{\rm Snoise}}
\newcommand {\ergcms}        {{\rm erg\,cm^{-2}\,s^{-1}}}
\newcommand {\ctsbeam}       {{\rm ct\,s^{-1}\,beam^{-1}}}
\newcommand {\SB}            {S_{\rm B}}
\newcommand {\ixrb}          {\langle I_{\rm XRB}\rangle}
\newcommand {\hmpc}          {h^{-1}\,{\rm Mpc}}

\newcommand{\ROSAT}{{\sl ROSAT}}
\newcommand{\Ginga}{{\sl Ginga}}
\newcommand{\ASCA}{{\sl ASCA}}

\begin{document}

\title[Soft X--ray background fluctuations and large scale structure in the
 Universe]
{Soft X--ray background fluctuations and large scale structure in the 
 Universe}

\author[F.J. Carrera, A.C. Fabian \& X. Barcons]
	{ F.J. Carrera$^{1,2}$, A.C. Fabian$^{3}$ and X. Barcons$^{1}$ 
	\\
	$^1$ Instituto de F\'{\i}sica de Cantabria (Consejo Superior
	de Investigaciones Cient\'{\i}ficas - Universidad\\ 
	de Cantabria), Avenida de los Castros, 39005 Santander, Spain\\
	$^2$ Mullard Space Science Laboratory--University College London,\\
	Holmbury St. Mary, Dorking, Surrey RH5 6NT, United Kingdom\\
	$^3$ Institute of Astronomy, Madingley Road,\\ Cambridge CB3 0HA,
	United Kingdom\\}

\maketitle

\begin{abstract}      

We have studied the fluctuations of the soft (0.9--2~keV) X--ray
background intensity for $\sim$10~arcmin and $\sim$2~arcmin beam
sizes, using 80 high galactic latitude medium--deep images from the
\ROSAT\ position sensitive proportional counter (PSPC). 
These fluctuations are dominated (and well reproduced) by confusion
noise produced by sources unresolved with the beam sizes we used. We
find no evidence for any excess fluctuations which could be attributed
to source clustering. The 95 per cent confidence upper limits on
excess fluctuations $\deltaiclus$ are: $(\deltaiclus/I)_{\rm
10~arcmin}\approxlt 0.12$, $(\deltaiclus/I)_{\rm 2~arcmin}\approxlt
0.07$.  We have checked the possibility that low surface brightness
extended objects (like groups or clusters of galaxies) may have a
significant contribution to excess fluctuations, finding that they are
not necessary to fit the distribution of fluctuations, and obtaining
an upper limit on the surface density for this type of source.
Standard Cold Dark Matter models would produce $\Delta I /I$ larger
than the above limits for any value of the density of the Universe
$\Omega=0.1-1$, unless the bias parameter of the X--ray emitting
matter is smaller than unity, or an important fraction of the sources
of the soft X--ray background ($\sim$30 per cent) is at redshifts
$z>1$.  Limits on the 2--10~keV excess fluctuations are also
considered, showing that X--ray sources in that band have to be at
redshifts $z>1$ unless $\Omega>0.4$.  Finally, if the spatial
correlation function of the sources that produce these excess
fluctuations is instead a power law, the density contrast $\droro$
implied by the excess fluctuations reveals that the Universe is smooth
and linear on scales of tens of Mpc, while it can be highly
non--linear on scales $\sim 1$~Mpc.

\end{abstract}


\section{INTRODUCTION}

Recent optical identification projects using \ROSAT\ PSPC observations have 
resolved an important fraction of the extragalactic X--ray Background (XRB) in
the $\sim$1--2~keV band into discrete sources, mostly Active Galactic Nuclei
(AGN) and Narrow Emission Line Galaxies (NELGs, a mixed bag including
Seyfert~2 galaxies, starburst galaxies and galaxies with HII regions)
(Page et al. 1996a, Jones et al. 1995, Boyle et al. 1995, Boyle et al. 1994).

Below 0.5~keV the fraction of the XRB that is extragalactic is uncertain, with
estimates ranging from about 10 to 20 per cent (McCammon \& Sanders 1990,
Barber \& Warwick 1994). The rest has a local origin, probably in a bubble of
hot gas surrounding the Sun. Above 2~keV, only $\sim$4 per cent of the XRB has
been resolved. Ongoing identifications of serendipitous sources in \ASCA\
images have increased this fraction to about 40 per cent (Inoue et al. 1996). 

Whatever the nature of the sources that produce the XRB, and independently of
their identification, the intensity of the XRB received from different
directions in the sky contains information on the angular distribution and
clustering properties of such sources. The study of the distribution of XRB
intensities $P(I)$ probes the source flux distribution ($dN/dS$ or number of
sources per sky area per unit flux) down to fluxes $S$ below the detection
limit (Barcons et al. 1994, Hasinger et al. 1993). This technique is called
$P(D)$, ($D=I-\ixrb$) or fluctuation analysis and is most sensitive to fluxes
in which there is about one source per `beam' (Scheuer 1974,  Barcons 1992),
the reason being that brighter sources contribute to the bright tail of the
(skewed) distribution, while fainter and more numerous sources produce
negligibly small noise. However, if the counting noise is important, the
technique is only sensitive to source fluxes equivalent to the photon
counting noise level.

The effect of source clustering is to decrease the effective number of
sources per beam, hence broadening $P(I)$ (Barcons 1992). This broadening
can be related to the  clustering properties of the sources that produce the
XRB, which in turn are due to density fluctuations in the Universe $\droro$
(Butcher et al. 1996, Barcons \& Fabian 1988, Rees 1980).

Instead of following the usual approach of using the deepest fields available
to push our knowledge of $dN/dS$ well below the present detection limits, in
this work we have explored the clustering properties of X--ray sources by
measuring or limiting the excess fluctuations they produce. Direct deep source
counts have been performed over small sky areas, and they might be biased by
large scale fluctuations in the source counts. The use of 80 widely scattered
\ROSAT\ fields allows a statistical study to be made (through
$P(I)$), avoiding any such biases.

The limits obtained on the excess fluctuations are then compared with the
specific expectations from  a Cold Dark Matter (CDM) model, to constrain the
density of the Universe ($\Omega\equiv 2q_0$) and/or the bias parameter of
X--ray emitting matter with respect to the underlying matter distribution 
($b_{\rm X}$). Assuming instead a power--law shape for the spatial correlation
function of the source of the soft XRB, the upper limits obtained on the excess
fluctuations have been used to investigate $\droro$ on different scales. 

In Section 2 we describe the data used in this work and the reduction process.
A brief summary of $P(I)$ analysis is given in Section 3, along with
the $dN/dS$ models used and the results of fitting the theoretical $P(I)$
curves to the data.

Section 4 is devoted to the development of the theoretical framework necessary
to relate these excess fluctuations to CDM power spectra and $\droro$ . The 
limits obtained on $\Omega$ and $b_{\rm X}$ are also presented and discussed,
as well as those obtained on $\droro$. In Section 5 we summarize our results.

We have parametrized the Hubble constant as $H_0=100\,h\,{\rm
km\,s^{-1}\,Mpc^{-1}}$, with $h=0.5$. The X--ray fluxes $S$ will be
given by default in the 0.5--2~keV range.

\section{THE DATA}

The data used in this work consist of 80 \ROSAT\ PSPC pointings with exposure
times longer than 8~ks at galactic latitudes higher than $20^\circ$. These same
fields were used for the RIXOS survey (Mason et al., in preparation). In
addition, the RIXOS fields were chosen avoiding extended or very bright targets
(e.g. clusters, nearby bright galaxies and bright stars). %#1

The Starlink software package ASTERIX was used for the data reduction.
The data were screened for high particle background intervals (Plucinsky
et al. 1993), bad aspect ratio solutions, and total accepted count rates
deviating from the average of each observation. This procedure normally
reduced the nominal exposure time by 10 to 20 per cent.
The remaining particle background was then calculated using the formulae in
Plucinsky et al. (1993), and subtracted. 

The remaining counts in Pulse Height Analyzer (PHA) channels 92 to 201
($\sim 0.7-2$~keV) for each pointing were then binned to obtain 
images with a pixel size of 4.5 arcsec. %#2
These images were then devignetted
by dividing by the exposure maps provided by the standard EXSAS
processing, after normalizing the maps to unity in the centre.
We note however that the results given below are practically insensitive  to
whether the remaining particle background is subtracted or not, or on whether
the vignetting has been corrected for or not.

The range of channels used in this work was chosen to avoid local contributions
to the XRB (such as the local bubble and Galactic diffuse emission, both
thought to be important only below $\sim$1~keV), solar contamination (usually
modelled as an oxygen line at about 0.5~keV, Snowden \& Freyberg 1993) and
absorption from neutral hydrogen (practically absent above 1~keV). An estimate
of the possible solar contamination was obtained by extracting images just
using night time observations (Snowden \& Freyberg 1993). This reduced
dramatically the  total number of counts, hence worsening the statistics,
without actually changing significantly the average count rate. We have,
therefore, used both day and night time data.

A circle of radius 5~arcmin around the target of each of the PSPC fields
(generally at the centre) was excluded. This proved sufficient to exclude
contributions from the targets down to the level in which their `tails' would
contribute less than 30 per cent of the local background per pixel in the two
worst cases. In most of them this contribution was $\approxlt5$ per cent.

%#1
Only one of the detected sources in the analyzed area in these fields is above
the flux interval used in our calculations (see Section 3.1), excluding that
field from our analysis does not affect any of our results, therefore we have
used the 80 fields including the detected sources within the regions explained
below.

Counts in each of the devignetted, particle--background subtracted,
target--subtracted images were further grouped in two beam sizes:

\begin{itemize}

\item An annulus of radii 5 and 10~arcmin centred on the pointing direction
(the inner radius is due to the target subtraction), giving a beam size of
$\Omegaeff=\pi\times(10^2-5^2)/3600= 0.06545\, {\rm deg^{-2}}$.  The
distribution, $P(I)$, of the 80 XRB intensities obtained (in counts per second
per `beam') ,$I$, is shown in Fig. 1.

\item Eight circles of radius 2.5~arcmin with centres equally spaced in
a circumference of radius 7.5~arcmin centred on the pointing
direction, hence $\Omegaoeff=\pi\times(2.5/60)^2=0.00545\, {\rm
deg^{-2}}$. We excluded two of these circular beams because more than
one third of their area was taken away by the target exclusion circle
(that was slightly off centre).  The remaining 638 values ($80\times
8\,-2$), $I_{\rm S}$, again in $\ctsbeam$, give $P(I_{\rm S})$, as
shown in Fig. 2.

\end{itemize}

\begin{figure}
  \vbox to 0cm{\vfil}
\epsfverbosetrue
\epsfysize= 180pt
\epsffile{fig1.vps}
\caption{Histogram of the distribution of the XRB intensities for the large
beam (see text). Also shown as a solid continuous line is the best fit $P(I)$
with $K=55$~deg$^{-2}$, with no cluster contribution and the average
$\deltainoise$
(see text).}
  \label{Fig1}
\end{figure}

\begin{figure}
  \vbox to 0cm{\vfil}
\epsfverbosetrue
\epsfysize= 180pt
\epsffile{fig2.vps}
\caption{Histogram of the distribution of XRB intensities for the small
beam (whole dataset, see text). Also shown as a solid continuous line is the
best fit $P(I_{\rm S})$ with $K=55$~deg$^{-2}$, with 
no cluster contribution and the
average $\deltaionoise$ (see text).}
  \label{Fig2}
\end{figure}

Both sets of intensities cover similar detector zones, but they sample
different angular scales: 10 to 15~arcmin in the first case and $<$5~arcmin in
the second. The maximum offaxis angle used (10~arcmin) ensures that the
vignetting correction is small ($<5$ per cent) and that the effective area is
also uniform over the detector region used.

We found average values of the XRB intensity of $0.49\pm0.02\,{\rm
ct\,s^{-1}\,deg^{-2}}$ from the large beam sample and $0.50\pm0.03\,{\rm
ct\,s^{-1}\,deg^{-2}}$ from the small beam sample (both 1 sigma confidence
intervals). We adopt $\ixrb=0.50\pm0.03\,{\rm ct\,s^{-1}\,deg^{-2}}$.

The  photon  counting noise was estimated by the square root of the number of
counts in each `beam' (using poisson statistics). We found $\deltainoise$ 
$=0.0018\pm0.0005$ $\,\ctsbeam$ and $\deltaionoise$ $=0.0005\pm0.0002$
$\,\ctsbeam$, in both cases we give 1 sigma confidence intervals.

A conversion factor of
$1\,{\rm ct\,s^{-1}}$~(92--201) = 
$2.02\times 10^{-11}\,{\rm erg\,cm^{-2}\,s^{-1}}$~(0.5--2~keV)
was used throughout, accurate within $\sim$5 per cent for power--law
energy spectral indices $\alpha\sim 0.4-0.7$, hydrogen column
densities $N_{\rm H}\sim (0.5-20)\times10^{20}$~cm$^{-2}$ and any
combination of detector response matrix and effective area, thus
covering the observed XRB spectrum (Gendreau et al. 1995,
Branduardi--Raymont et al. 1994) and the galactic columns of the \ROSAT\
observations used (Mason et al. 1996).

We therefore measure a total XRB intensity (including sources) of
$\ixrb=(3.3\pm0.3)\times 10^{-8}\,{\rm erg\,cm^{-2}\,s^{-1}\,sr^{-1}}$
(0.5--2~keV). This value is somewhat higher than previous XRB intensity
estimates, but still overlaps within $\sim$2 sigma with the value  obtained by
Branduardi--Raymont et al. (1994), for example.

\section{FLUCTUATION ANALYSIS}

\subsection{Contribution from point sources}

In this work we have adopted the $dN/dS$ shape and
parameters from Barcons et al. (1994):

\[
{dN\over dS}(S)={K\over \SB}\,
   \left({S\over \SB}\right)^{-\gamma_{\rm d}}\,\,\, S<\SB
\]

\[
{dN\over dS}(S)={K\over \SB}\,
   \left({S\over \SB}\right)^{-\gamma_{\rm u}}\,\,\, S>\SB %#3
\]

\noindent with $\SB=2.2\times 10^{-14}\,\ergcms$, $\gamma_{\rm
d}=1.8$,  $\gamma_{\rm u}=2.5$\ and $K=$55~deg$^{-2}$.

The results given below do not change if we use the slightly different
parameters from Branduardi--Raymont et al. (1994) or Hasinger et al. (1993),
which is hardly surprising considering that they are all mutually consistent,
have been obtained with \ROSAT\ data and sample similar or overlapping flux
ranges. This also means that no biases have been introduced in the
determination of the source counts in those surveys by large scale source
number fluctuations.

%#1

The $dN/dS$ parameters given above are appropriate for $S$ between 0.07 and
50$\times 10^{-14} \,\ergcms$. At the level of one source per beam, the $P(I)$
curve is going to be sensitive down to fluxes $S\sim 4\times 10^{-14}
\,\ergcms$ and the $P(I_{\rm S})$ down to $S\sim 0.5\times 10^{-14} \,\ergcms$.
The sensitivity limit of our analysis is thus $S\sim10^{-14} \,\ergcms$. The
width of the $P(I)$ is mainly due to this `confusion noise' rather than to
photon counting noise.
Although we
are integrating the $dN/dS$ between zero and infinity in our calculations, the
practicalities of using a Fast Fourier Transform algorithm to calculate the
$P(I)$ effectively reduced this interval to
$(0.02-40)\times 10^{-14}\,\ergcms$.

Given a $dN/dS$ and a beam profile, the shape of $P(I)$ can be
predicted (see Barcons 1992 and references therein). The counting
noise is generally taken into account by convolving $P(I)$ with a
gaussian of width $\deltainoise$. We have also followed this approach,
taking as $\deltainoise$ the average values given above, and checking
the influence of the dispersion around those values by using the 1
sigma upper and lower limits as well (see below).

%
% essentially assuming
%that the number of counts is large enough for the gaussian and poisson
%distributions to be equivalent. This is the case for all our $I$ and
%$\sim$90 per cent of our $I_{\rm S}$, and consequently 
%

The beam functions are taken as two circular step functions: one with
an outer radius of 10~arcmin and an inner radius of 5~arcmin (for
$I$), and another with just an outer radius of 2.5~arcmin (for $I_{\rm S}$).
The sizes are much larger than the Point Spread Function (PSF) of
the XRT/PSPC combination (Hasinger et al. 1992), making the
convolution of the PSF and the step functions indistinguishable from
the simple step functions in practice.

Any width in excess of that expected from the source flux distribution and the
poisson counting noise is called excess variance, and is usually modelled by
convolving  $P(I)$ with a gaussian of width $\deltaiclus$.
We assume that the excess fluctuations arise from clustering of sources,
and perhaps some contribution from extended sources like clusters of galaxies
(see below). If any other unknown systematic effect contributes to the excess
fluctuations, the results given below would just be upper limits to
$\deltaiclus$ really due to clustering, and any consequences of the results
given here would be strengthened.

The model $P(I)$ is then (see also Eq. 22 in Barcons 1992):

\[
P(I)=\int d\omega\,  e^{-2\pi i\omega I} \exp\left\{ 
     -\omega^2\deltainoise^2/2-\omega^2\deltaiclus^2/2 \right\}
\]

\begin{equation}\label{PI}
\qquad\times\exp\left\{\Omegaeff \int dS dN/dS  
                 \bigl[ \exp (2\pi i\omega S/\Omegaeff)-1  \bigr]\right\}
\end{equation}

The same expression is valid for $I_{\rm S}$ replacing $\Omegaeff$ with
$\Omegaoeff$.

\subsection{Contribution from extended sources}

We have considered the X--ray emitting clusters reported by Rosati et al.
(1995). With the above parametrization, a set of parameters that follow their
$dN/dS$ in the flux range $(1-40)\times 10^{-14}\,\ergcms$ is:
$\gamma_d=\gamma_u=1.962$, $K_{\rm cl}=9.784$~deg$^{-2}$\ and 
$S_{\rm Bcl}=10^{-14}\,\ergcms$.

The properties of these clusters have been taken from the study of poor groups
of galaxies by Mulchaey et al. (1996). We have assumed the temperature of the
hot gas (responsible for the detected X--ray emission) to be $kT\sim1$~keV and
a King emission profile with a cluster core size  of $R_{\rm core}=15$~arcmin
(changing the size to 7~arcmin does not affect the results given below). For a
nearby group (like those in Mulchaey et al. 1996) with $z\sim0.02$ this
corresponds to a core size of $\sim0.4$~Mpc (or $\sim0.2$ for 7~arcmin).

The conversion factor for clusters, assuming a thermal bremstrahlung spectrum
with the above temperature and absorption by neutral hydrogen with
$N_{\rm H}=10^{20}$~cm$^{-2}$, is
$1\,{\rm ct\,s^{-1}}$ (92--201) = 
$1.64\times 10^{-11}\, {\rm erg\,cm^{-2}\,s^{-1}}$ (0.5--2~keV).

The cluster contribution to $P(I)$ is modelled by convolving it with
the $P(I)$ due to the clusters only, i.e. by adding another term to the
exponent in braces in Eq. \ref{PI}

\[
P(I)=\int d\omega\,  e^{-2\pi i\omega I} \exp\left\{ 
     -\omega^2\deltainoise^2/2-\omega^2\deltaiclus^2/2 \right\}
\]
\[
\quad\times\exp\left\{\Omegaeff \int dS dN/dS  
                 \bigl[ \exp (2\pi i\omega S/\Omegaeff)-1 
                 \bigr]\right\}
\]
\begin{equation}\label{PIclus}
\quad\times\exp\left\{
2\pi\!\!\!\int\!\!\!dr r\!\!\!\int\!\!\!dS
(dN/dS)_{\rm cl}\bigl[ \exp(2\pi i\omega S G_{\rm cl}(r))-1\! \bigr]
\right\}
\end{equation}

\noindent where $G_{\rm cl}(r)$ is the convolution of a King profile with
the step functions described above.

Only clusters with fluxes $S>10^{-14}\,\ergcms$ (the sensitivity limit of the
Rosati et al. sample) have been used to calculate the $P(I)$. Again, the results
given below do not change if we decrease this limit by a decade, because
of the flatness of the clusters source counts.

The angular size of the clusters of galaxies considered here implies
that, if one of them is present in a given \ROSAT\ pointing, the eight
small beams will be affected. This introduces a correlation between
them and complicates the error estimates on $\deltaioclus$. A way
around this problem is to select one of the eight beams for each
\ROSAT\ pointing at random, and just use those 80 values of $I_{\rm
S}$. This allows us to estimate the significance of the cluster
contribution (at the price of sacrificing sensitivity). Should this
contribution prove to be negligible, the whole dataset can be used,
applying Eq. \ref{PI} instead of Eq. \ref{PIclus}.

\subsection{Fitting process and results}

\begin{figure}
  \vbox to 0cm{\vfil}
\epsfverbosetrue
\epsfysize= 180pt
\epsffile{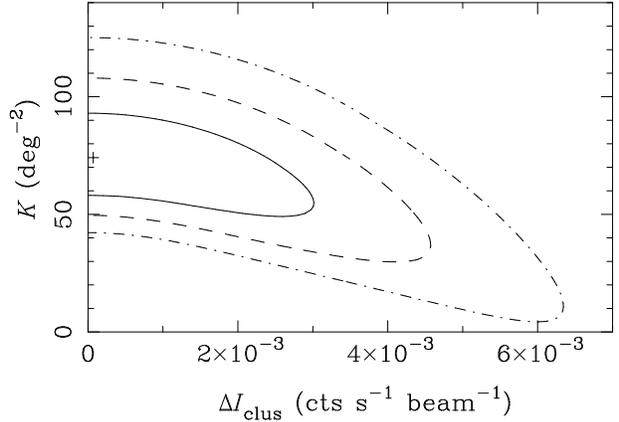}
\caption{Contours of $\Delta$L values (1, 2 and 3 sigma) in
($K$,$\deltaiclus$) space, for the large beam, with no cluster contribution
and the mean $\deltainoise$ (see text)}
  \label{Fig3}
\end{figure}

\begin{figure}
  \vbox to 0cm{\vfil}
\epsfverbosetrue
\epsfysize= 180pt
\epsffile{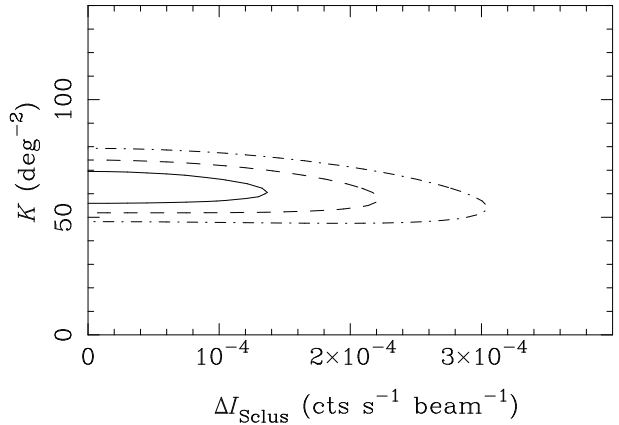}
\caption{Contours of $\Delta$L values (1, 2 and 3 sigma) in
($K$,$\deltaioclus$) space, for the small beam, with no cluster contribution,
the mean $\deltaionoise$ and the whole dataset (see text)}
  \label{Fig4}
\end{figure}

A Maximum Likelihood fitting method was adopted. $\chi^2$ was not adequate
because the number of fitting points for $P(I)$ was too small to make a
significant number of bins with a reasonable number of points in each one of
them (enough for gaussian statistics to be valid).

A further ingredient (apart from $dN/dS$, $(dN/dS)_{\rm cl}$,
$\deltainoise$\, and $\deltaiclus$) is necessary to fit $P(I)$: the total
intensity of the sources in the $dN/dS$ used to calculate $P(I)$  ($\langle I
\rangle _{dN/dS}$) is always smaller than the mean observed intensity. The
missing sources are {\it not} important for the shape of $P(I)$, because
there are so many of them and  they are so faint that they contribute a
negligibly small gaussian noise to it (already taken into account with
$\deltainoise$). Their absence makes the `peak' of the model $P(I)$ to be 
at an intensity smaller than that of the peak of the observed distribution, %#
so an overall shift of the distribution is
necessary to compare the observed and modelled $P(I)$.

An additional intensity $\Delta I$ is added to each $I$ to shift them to higher
values and is allowed to vary until a best fit is obtained (keeping the rest of
the parameters fixed). It is then discarded as a non--interesting parameter and
the fitting proceeds with a different set of parameters. The best fit $\Delta
I$ is in fact only very weakly dependent on the rest of the parameters. The
value of $\Delta I$ can however be predicted from the $dN/dS$ and $\langle
I_{\rm XRB}\rangle$, and its final best fit value is not expected to be very
different from this predicted value.

For each set of fitting parameters, we have defined the likelihood function as

\begin{eqnarray}
{\rm L}(\deltaiclus,K)&=&-2\sum_i\ln P(I_i)\nonumber\\ 
&+& \left( 
     {\ixrb -\Delta I -\langle I\rangle_{dN/dS} \over \Delta\ixrb}
  \right)^2
\end{eqnarray}

\noindent where the first term is the usual definition
(and $P(I)$ is as defined in Eq. \ref{PI} or \ref{PIclus}), %#5
and the second term
makes added intensities far from their expected values less likely, weighted
for each beam size with the error in the estimate of the XRB intensity,
$\Delta\ixrb$, given above. With this
definition $\Delta$L is distributed as $\Delta\chi^2$.

\begin{table*}
\centering
  \vbox to 0cm{\vfil} 
\begin{minipage}{22cm} 
\caption{Results of the fit to $P(I)$.}
 \label{Table1} 
\begin{tabular}{ c c c c c c c }
$\deltainoise$ & $\deltaiclus$ & 2$\sigma$ upper limit & L & $K$           &
$K_{\rm cl}$ & 2$\sigma$ upper limit \\
($\ctsbeam$)   & ($\ctsbeam$)  & ($\ctsbeam$)          &   & (deg$^{-2}$)  &
(deg$^{-2}$) & (deg$^{-2}$)          \\
\hline
Mean           & 0.0017  & 0.0037 & -556.3 & 55.0 fixed  & 9.8 fixed & -    \\
Mean$+1\sigma$ & 0.0008  & 0.0034 & -556.3 & "           & "         & -    \\
\smallskip
Mean$-1\sigma$ & 0.0021  & 0.0039 & -556.3 & "           & "         & -    \\

Mean           & 0.0006  & 0.0038 & -556.8 & 55.0 fixed  & 25.8      & 62.2 \\
Mean$+1\sigma$ & 0.0000  & 0.0035 & -556.6 & "           & 19.3      & 55.8 \\
\smallskip
Mean$-1\sigma$ & 0.0013  & 0.0039 & -556.8 & "           & 26.0      & 67.2 \\

Mean           & 0.0001  & 0.0038 & -557.0 & 67.3        & 9.8 fixed & -    \\
Mean$+1\sigma$ & 0.0000  & 0.0037 & -556.7 & 61.9        & "         & -    \\
\smallskip
Mean$-1\sigma$ & 0.0003  & 0.0040 & -557.1 & 71.1        & "         & -    \\

Mean           & 0.0022  & 0.0041 & -555.6 & 55.0 fixed  & 0.0 fixed & -    \\
Mean$+1\sigma$ & 0.0016  & 0.0038 & -555.6 & "           & "         & -    \\
\smallskip
Mean$-1\sigma$ & 0.0025  & 0.0043 & -555.6 & "           & "         & -    \\

Mean           & 0.0001  & 0.0038 & -557.1 & 74.1        & 0.0 fixed & -    \\
Mean$+1\sigma$ & 0.0003  & 0.0037 & -556.5 & 62.9        & "         & -    \\
Mean$-1\sigma$ & 0.0001  & 0.0039 & -557.2 & 78.3        & "         & -    \\
\hline
   \end{tabular}
 \end{minipage}
\end{table*}

\begin{table*}
\centering
  \vbox to 0cm{\vfil} 
\begin{minipage}{22cm} 
\caption{Results of the fit to $P(I_{\rm S})$.}
 \label{Table2} 
\begin{tabular}{ c c c c c c c c }
$N$&$\deltaionoise$ &$\deltaioclus$ & 2$\sigma$ upper limit & L & $K$         &
     $K_{\rm cl}$ & 2$\sigma$ upper limit\\
   &($\ctsbeam$)    &($\ctsbeam$)   & ($\ctsbeam$)          &   & (deg$^{-2}$)&
     (deg$^{-2}$) & (deg$^{-2}$)         \\
\hline
80 & Mean           & 0.0000 & 0.0004 & -861.7 & 55.0 fixed & 9.8 fixed & -  \\
"  & Mean+1$\sigma$ & 0.0000 & 0.0004 & -857.5 & "          & "         & -  \\
\smallskip
"  & Mean-1$\sigma$ & 0.0003 & 0.0005 & -861.9 & "          & "         & -  \\

80 & Mean           & 0.0000 & 0.0004 & -862.1 & 55.0 fixed & 0.0       &49.6\\
"  & Mean+1$\sigma$ & 0.0000 & 0.0004 & -858.6 & "          & 0.0       &31.9\\
\smallskip
"  & Mean-1$\sigma$ & 0.0003 & 0.0006 & -862.2 & "          & 0.0       &83.9\\

80 & Mean           & 0.0000 & 0.0007 & -862.0 & 49.5       & 9.8 fixed & -  \\
"  & Mean+1$\sigma$ & 0.0000 & 0.0006 & -861.8 & 31.0       & "         & -  \\
\smallskip
"  & Mean-1$\sigma$ & 0.0003 & 0.0008 & -861.9 & 55.0       & "         & -  \\

80 & Mean           & 0.0000 & 0.0004 & -862.1 & 55.0 fixed & 0.0 fixed & -  \\
"  & Mean+1$\sigma$ & 0.0000 & 0.0004 & -858.6 & "          & "         & -  \\
\smallskip
"  & Mean-1$\sigma$ & 0.0003 & 0.0006 & -862.2 & "          & "         & -  \\

80 & Mean           & 0.0000 & 0.0007 & -862.2 & 52.5       & 0.0 fixed & -  \\
"  & Mean+1$\sigma$ & 0.0000 & 0.0006 & -862.0 & 34.1       & "         & -  \\
"  & Mean-1$\sigma$ & 0.0003 & 0.0008 & -862.2 & 55.0       & "         & -  \\
\hline
638& Mean           & 0.0000 & 0.0002 & -6820  & 55.0 fixed & 0.0 fixed & -  \\
 " & Mean+1$\sigma$ & 0.0000 & 0.0002 & -6797  & "          & "         & -  \\
\smallskip
 " & Mean-1$\sigma$ & 0.0003 & 0.0004 & -6820  & "          & "         & -  \\

638& Mean           & 0.0000 & 0.0002 & -6823  & 62.6       & 0.0 fixed & -  \\
 " & Mean+1$\sigma$ & 0.0000 & 0.0002 & -6800  & 47.3       & "         & -  \\
 " & Mean-1$\sigma$ & 0.0000 & 0.0002 & -6836  & 76.5       & "         & -  \\
\hline
   \end{tabular}
 \end{minipage}
\end{table*}

The first fit is performed fixing all the $dN/dS$ parameters to the values
given above and leaving $\deltaiclus$ as the only free parameter. The best fit
values are shown in Table 1 (for the large beam) and Table 2 (for the small
beam). The effect of the uncertainty on $\deltainoise$ has been assessed by
fixing it to its mean value and the 1 sigma upper and lower limits, and
performing the fit for each of these three values. The results are indicated in
Tables 1 and 2 (rows with both the $K$ and $K_{\rm cl}$ columns labelled
`fixed'), with the first row of each group of three corresponding to the mean,
and the second and the third line to the 1 sigma upper limit and lower limit,
respectively. At the 2 sigma confidence level, only upper limits are obtained:
$\deltaiclus<0.004\,\ctsbeam$ and $\deltaioclus<0.0005\,\ctsbeam$ (or
$\deltaiclus/\ixrb<$12 per cent and $\deltaioclus/\ixrb<$19 per cent).

The $dN/dS$ normalization, $K$, and $\deltaiclus$ are coupled to some extent:
large normalizations increase the `intrinsic' $P(I)$ width, thus reducing the
amount of excess variance needed. We have done a second set of fits in two
dimensions, with both $K$ and $\deltaiclus$ as free parameters. The results are
shown in Tables 1 and 2. $\Delta$L contours are plotted in  Fig. 3 (large beam)
and 4 (small beam), for the case of no cluster contribution (and the whole
dataset, see below) and the average values of $\deltainoise$ and
$\deltaionoise$, respectively.

It is possible to obtain confidence intervals on $\deltaiclus$ taking into
account its coupling with $K$ by finding the minimum $\Delta$L value as a
function of $K$ for every $\deltaiclus$, and then considering them as a one
dimensional $\Delta$L profile for $\deltaiclus$ (Lampton, Margon and Bowyer
1976). This has been done for the results plotted in Fig. 3 and 4, and it is
shown in Fig. 5 and 6, respectively, as well as in Tables 1 and 2. As in the one
dimensional case, at the 2 sigma confidence level, only upper limits are
obtained: $\deltaiclus<0.004\,\ctsbeam$ and
$\deltaioclus<0.0006-0.0008\,\ctsbeam$ (or $\deltaiclus/\ixrb<$12 per cent and
$\deltaioclus/\ixrb<$22--30 per cent).

\begin{figure}
  \vbox to 0cm{\vfil}
\epsfverbosetrue
\epsfysize= 180pt
\epsffile{fig5.vps}
\caption{One dimensional $\Delta$L profiles extracted from the contours
in Fig. 3 (see text) as a function of $\deltaiclus$. The solid line
corresponds to the mean $\deltainoise$, and the dashed and dotted lines
correspond to adding and subtracting 1$\sigma$ from it, respectively.}
  \label{Fig5}
\end{figure}

\begin{figure}
  \vbox to 0cm{\vfil}
\epsfverbosetrue
\epsfysize= 180pt
\epsffile{fig6.vps}
\caption{One dimensional $\Delta$L profiles extracted from the contours
in Fig. 4 (see text) as a function of $\deltaioclus$. The solid line
corresponds to the mean $\deltaionoise$, and the dashed and dotted lines
correspond to adding and subtracting 1$\sigma$ from it, respectively.}
  \label{Fig6}
\end{figure}

All the above fits have been repeated without any cluster contribution, and the
results also included in Tables 1 and 2. It is clear that adding the clusters
does not significantly reduce  the L values, nor does it reduce the excess
variance. We obtained a quantitative assessment of the significance of this
contribution using the standard F--test (Bevington 1969). This assesses the
relative improvement in $\chi^2$ (or L) on the addition of a new free fitting
parameter ($K_{\rm cl}$); in our case that means comparing the values of L in
the second and fourth groups of rows in Tables \ref{Table1} and \ref{Table2}.
The F--test reveals that the addition of $K_{\rm cl}$ does not improve
significantly the fits, to a confidence of 96 per cent for the large beam, and
the best fit for the small beam actually corresponds to $K_{\rm cl}=0$. We can
then conclude that their contribution to the $P(I)$ width is negligible and
ignore it. This allows us to use the full 638 values of $I_{\rm S}$, reducing
considerably the 2 sigma upper limit in the excess variance:
$\deltaioclus<0.0002\,\ctsbeam$ (or $\deltaioclus/\ixrb<$7 per cent).

Confidence regions on $K_{\rm cl}$ can be obtained from the $\Delta$L contours
in the $(\deltaiclus,K_{\rm cl})$ space with the method described above. Only
upper limits are obtained at 2 sigma level, and are given in Tables
\ref{Table1} and \ref{Table2}. Rosati et al. warn that their value is only a
lower limit to the real surface density of clusters (or extended X--ray
sources). Our results show that down to $10^{-15}-10^{-14}\,\ergcms$, the
surface density of clusters is not larger than 3 to 6 times the value obtained
by Rosati et al.

%#4

So\l tan et al. (1996) found an important contribution ($\sim$30 per
cent) to the angular correlation function of the soft XRB from
extended haloes around Abell clusters of galaxies on scales
$>1$~degree. Since we are exploring much smaller angular scales and
the opposite (low flux) end of the $dN/dS$ distribution of the X--ray
emitting clusters, there is no contradiction between our finding that
extended sources (clusters) do not contribute significantly to the
excess fluctuations and the results of So\l tan et al. (1996).

The upper limits on the excess fluctuations obtained in this section (namely,
$\deltaiclus <$12 per cent and $\deltaioclus <$7 per cent, with a 2 sigma
confidence level), will be used in Section 5 to constrain the values of the
density parameter of $\Omega$ and $b_{\rm X}$ using the expressions derived in
Section 4.

\section{Inhomogeneities  in the mass distribution of the Universe}

\subsection{Relation of excess fluctuations to the power spectrum}

It is easy to realize that $(\deltaiclus/\ixrb)^2$ is the value of the
autocorrelation function of the XRB at zero--lag. We can then
use the expressions in Appendix A of Barcons \&
Fabian (1988) and Eqs. 2 and 4 of Carrera et al. (1991) to relate the
limits found on the excess fluctuations to the clustering properties of
the sources of the XRB.

In our case, the beam shape is a two sided step function, with a value of 1
between $r_1$ and $r_2$ and 0 outside, where $r_1=5$~arcmin and $r_2=10$~arcmin
for the large beam, and $r_1=0$ and $r_2=2.5$~arcmin for the small beam. Its
two dimensional Fourier transform is

\begin{equation}
\hat G(q)=(r_2 J_1(r_2 q) -r_1 J_1(r_1 q))/q
\end{equation}

\noindent where $q$ is the magnitude of the two dimensional Fourier
space vector, and $J_1(x)$ is the Bessel function of order 1.

Solving Eq. 2 and 4 in Carrera et al. (1991) for $\ixrb$, we arrive at

\[
{1\over f^2}\left({\deltaiclus\over \ixrb}\right)^2={1\over 4\sqrt{2\pi}}
\]
\[
\qquad\times{c\over H_0} \int dz\, (1+z)^{-8}(1+2q_0z)^{-1/2} 
              j^2(z)/d_{\rm A}^2(z)
\]
\[
\qquad\qquad\int d^2q\,\hat G^2(q)\hat\xi(q/d_{\rm A}(z))
\]
\begin{equation}\label{dii}
\qquad\times\left[ {\Omegaeff\over4\pi}
  {c\over H_0}\int dz\, (1+z)^{-5} (1+2q_0z)^{-1/2} j(z)  \right]
  ^ {-2}
\end{equation}

\noindent $d_{\rm A}(z)$ being the angular distance and $j(z)$ the K--corrected
%#6
volume emissivity (emitted power per unit volume) of the sources that produce
the excess variance and contribute a fraction $f$ to the XRB. $\hat\xi(k)$ is
the three dimensional Fourier transform of the spatial correlation function
$\xi(r)$, and $k$ is the magnitude of the three dimensional Fourier space
vector. Following Peebles (1980), $\hat\xi(k)$ is also the power spectrum,
multiplied by $(2/\pi)^{3/2}$, due to the different definitions of the Fourier
transform used here and in Peebles (1980).

Eq. \ref{dii} allows the calculation of $\deltaiclus/\ixrb$ for a particular
$b_{\rm X}$ and a power spectrum model, which in turn would depend on $\Omega$
(see below). By comparing these predictions with the upper limits obtained
above, constraints can be placed on those cosmological parameters. In the next
section we present the luminosity function we have used to calculate $j(z)$.

\subsection{Luminosity functions and modelling}

At the flux levels at which our $P(I)$ anaysis is sensitive ($S\sim
10^{-14}\,\ergcms$) the dominant type of X--ray sources found in
\ROSAT\ surveys are AGN although with an increasingly important
fraction of NELGs (McHardy et al., in preparation, Mason et al., in
preparation, Boyle et al. 1995, Carballo et al. 1995, Boyle et
al. 1994).

The X--ray Luminosity Function (XLF, number of sources per unit volume and unit
luminosity) of AGN has been very well studied recently with \ROSAT\ at those
fluxes (Page et al. 1996a, Boyle et al. 1994). It has been found to be well
represented by a broken power law. Within a pure luminosity evolution model,
the AGN luminosities  have a fast positive evolution up to $z\sim1.5-2$. At
that redshift the evolution slows down, or even stops and becomes negative.

We have obtained the emissivity $j(z)$ in Eq. \ref{dii} by integrating the best
fit XLF models of Page et al. (1996a), since AGN are the main contributors to
the XRB over the flux range studied. Indeed, AGN are about $\sim$50 per cent of
the sources at the fluxes we are dealing with, and we have to consider the
redshift evolution of the volume emissivity from other sources. The evolution
of NELGs, the other type of source with a sizeable contribution and likely to
be clustered, is somewhat different (Page et al. 1996b, Boyle
et al. 1995). Their rate of evolution is lower than that of AGN, and they are
concentrated at low $z$ ($<0.6$).

We have adopted the best power--law model with cut off evolution and $q_0=0.5$
with a conversion factor of 1.8 (between \ROSAT\ and {\sl Einstein}\ fluxes)
from Page et al. (1996), but making $q_0$ half the value of $\Omega$
investigated in each case. The other best fit models produce very similar
$\deltaiclus/\ixrb$ values. The K correction has been calculated using
$\alpha\sim1$, as observed for AGN, the dominant type of sources in our flux
range (Mittaz et al., in preparation, Almaini et al. 1996, Ciliegi et al.
1996, Romero--Colmenero et al. 1996, Vikhlinin et al. 1995).

%#7

We have also considered the results on 2--10~keV \Ginga\ excess
fluctuations from Butcher et al.  (1996): $(\deltaiclus/\ixrb)<0.038$
(2 sigma). In this case the redshift dependence of the emissivity
$j(z)$ of the sources is not known and we have adopted a very simple
model for their redshift distribution: $j(z)\propto(1+z)^{3+p}$. $p=0$
corresponds to no evolution of the emissivity in comoving
coordinates. For a simple power law luminosity function, $p\sim3$
implies a luminosity evolution similar to that found in the soft
band. We have approximated the \Ginga\ collimator shape by a gaussian
of dispersion $\gamma_{\rm S}\sim0.8\,{\rm deg}$, and used an energy
index of $\alpha=0.7$ as observed for AGN in that band.

%
%a step function of width $\Delta z=1$ centred at $z=1$ (Eqs. A22 and A23 in
%Barcons \& Fabian 1988, with a beam size $\gamma_{\rm S}\sim0.8\,{\rm deg}$
%appropriate for \Ginga). The resulting $\deltaiclus/\ixrb$ limits are only
%weakly dependent on whether this step is placed at $z=1$ or 2. The  second
%value would be preferred by models which produce the XRB spectrum by
%superposition of AGN spectra (see e.g. Fabian \& Barcons 1992 and references
%therein).
%

Making $f=1$ in Eq. \ref{dii} is equivalent to assuming %#8
that the sources whose clustering produces the excess fluctuations we are
studying produce all the XRB. We know that only 50--60 per cent of the XRB 
is produced by sources with fluxes larger than $\sim 10^{-14}\,\ergcms$
(our sensitivity limit). However, since in Eq. \ref{dii} the absolute 
normalization of the XLF cancels out, just having more sources with
the same evolution would not affect our theoretical $\Delta I/I$. We have also
checked that extending the integrals in redshift in Eq. \ref{dii}
to $z=5$ instead of $z=3$ (our default value) does not affect our results.
If, as discussed above, the NELGs are proved to make an important contribution
to the XRB, but with a different evolution (more concentrated at lower $z$),
the resulting density fluctuations produced by these sources would be
larger, hence {\it strengthening} our results.

A similar argument can be used for the \Ginga\ upper limits.

\subsection{CDM Models}

Cold Dark Matter models present a picture of the Universe in which the smallest
structures (galaxies) form first and, by merging, form larger structures
(Peacock \& Dodds 1994). Even if the basic assumptions have not been
thoroughly tested, the CDM scenario provides useful calculation tools and
expressions to analyze the evolution of the Universe.

This is the case for the power spectrum of density fluctuations $P(k)$. A
number of useful parametrizations that fit some of the available angular  and
spatial clustering data are found in the literature (Peacock \& Dodds 1994
--hereafter PD--, Efstathiou, Bond and White 1992 --hereafter EBW--, Bardeen et
al. 1986).

We have used the shape of the {\it linear} power spectrum of PD:

\[
P(k)\propto {k^4\over 4\pi k^3}
\left\{
{ \ln(1 + gk) \over gk }\right\}^2
\]
\begin{equation} \label{CDM}
\qquad\times \left\{\left[ 1+ak+(bk)^2+(ck)^3+(dk)^4 \right]^{-1/4} 
\right\}^2
\end{equation}

\noindent where $a=(3.89/\Gamma) \,\hmpc$, $b=(14.1/\Gamma) \,\hmpc$,
$c=(5.46/\Gamma) \,\hmpc$, $d=(6.71/\Gamma) \,\hmpc$, $g=(2.34/\Gamma) \,\hmpc$,
and $\Gamma$ is a shape parameter that can be changed, both to make Eq.
\ref{CDM} fit several different observations, and to reflect the behaviour of
different CDM and Mixed Dark Matter models. Following PD, we have chosen

\begin{equation} \label{Gamma}
\Gamma=\Omega h \exp(-2\OmegaB)
\end{equation}

\noindent which is equivalent to that presented by EBW for zero baryonic
density $\OmegaB=0$, but also includes an empirical dependence in
$\OmegaB$, making high baryonic content models mimic low CDM
density. The power spectrum parametrization with the shapes and
parameters from EBW is similar (for a power spectrum index $n=1$).

%
%The range in wavenumber $k$ in which the above parametrizations are valid
%($0.01\approxlt k\approxlt1\,(\hmpc)^{-1}$) corresponds to the spatial scales 
%sampled here (see below), meaning that the results given below are not strongly
%dependent on the behaviour of the tails of the power spectrum.
%

\begin{table}
\centering
  \vbox to 0cm{\vfil}  
\caption{$\deltaiclus/\ixrb$ from Cold Dark Matter for \ROSAT.
Linear power spectrum from Peacock \& Dodds (1994)}
 \label{Table3} 
\begin{tabular}{ c c c c }
\hline
$\Omega$ & ${\deltaioclus\over\ixrb}$ & ${\deltaiclus\over\ixrb}$ \\  
\hline
         0.1 & 0.107 & 0.098 \\
         0.2 & 0.098 & 0.086 \\
         0.4 & 0.097 & 0.081 \\
         0.6 & 0.101 & 0.080 \\
         0.8 & 0.106 & 0.081 \\
         1.0 & 0.111 & 0.083 \\
\hline
Upper limits & 0.072 & 0.119 \\
\hline
   \end{tabular}
\end{table}

\begin{table}
\centering
  \vbox to 0cm{\vfil}  
\caption{$\deltaiclus/\ixrb$ from Cold Dark Matter for Ginga.
Linear power spectrum from Peacock \& Dodds (1994)}
 \label{Table4} 
\begin{tabular}{ c c c c c c }
\hline
 & \multicolumn{2}{c}{$p=0$} & & \multicolumn{2}{c}{$p=3$} \\
          \cline{2-3}                  \cline{5-6}         \\
$z_{\rm max}$ & 1     & 3    & & 1     & 3     \\
\hline 
$\Omega$      & \multicolumn{5}{c}{${\deltaiclus/\ixrb}$}  \\
\hline
         0.1  & 0.219 & 0.222 & & 0.066 & 0.113  \\
         0.2  & 0.174 & 0.176 & & 0.051 & 0.081  \\
         0.4  & 0.142 & 0.143 & & 0.040 & 0.059  \\
         0.6  & 0.129 & 0.129 & & 0.036 & 0.050  \\
         0.8  & 0.121 & 0.122 & & 0.033 & 0.045  \\
         1.0  & 0.116 & 0.117 & & 0.032 & 0.043  \\
\hline
Upper limit   & 0.038 & 0.038 & & 0.038 & 0.038  \\
\hline
   \end{tabular}
\end{table}

PD also give a dependency of the normalization of $P(k)$ with $\Omega$:
$\sigma_8=0.75 \Omega^{-0.15}$, $\sigma_8$ being the rms density contrast when
averaged over spheres of radius 8~$\hmpc$. We have adopted this normalization
dependence on $\Omega$. For each value of $\Omega$ we have calculated
$\sigma_8$ from Eq. \ref{CDM}, rescaling its normalization to give the value of
$\sigma_8(\Omega)$ given above. This normalized $P(k)$ is then used to
calculate $\Delta I/I$.

Standard primordial nucleosynthesis and abundances observations constrain
$\OmegaB\sim0.05$ (Olive \& Steigman, 1995), and we have assumed this value.
Since $\OmegaB$ only appears in an exponent and is small in any case, changing
it by $\pm 0.01$ (its observational confidence interval) does not change the
results given below.

X--ray sources are possibly more clustered than %#9
the underlying matter, and
therefore the $\deltaiclus/\ixrb$ obtained from CDM  has to be multiplied by
the bias parameter $b_{\rm X}$. A value $b_{\rm X}\sim (3.4\pm
0.8)\Omega^{0.6}/f'$ has been found for nearby bright X--ray sources, where
$f'$ is the fraction of the gravitational acceleration on the Local Group
contributed by the $z<0.015$ region ($f'\sim0.5$, Miyaji 1994). %#10

With \hfill all \hfill the \hfill above \hfill assumptions, \hfill our
\hfill CDM\\
$\deltaiclus / \ixrb$ only has
two free
parameters: $\Omega$ and $b_{\rm X}$. We have sampled $\Omega$ between 0.1 and
1, and assumed $b_{\rm X}=1$.
Different values of $\Omega$ change the shape of the CDM power spectrum
(through the shape parameter $\Gamma$), while the effect of $b_{\rm X}$ is just
multiplicative.

The above CDM power spectrum shape is constant in comoving coordinates. Its
evolution with redshift is obtained by multiplying its normalization by
a factor $D^2(z)$ that is proportional to $(1+z)^{-2}$ for $\Omega=1$, and has
a more complicated dependence with redshift for smaller values of $\Omega$
(Peebles 1980).

We present in Table \ref{Table3}  $\deltaiclus/\ixrb$ obtained for the beam
sizes and shapes used here (small and large beam), for several different values
of $\Omega$ in the above range, $b_{\rm X}=1$ and the PD power spectrum given
in Eq. \ref{CDM} (the power spectrum of EBW produces similar results). 
$\Delta I/I$ produced by CDM  exceeds our small beam upper limits
($\deltaioclus/\ixrb<0.07$,) for any value of $\Omega$. The power spectrum of
the spatial distribution of the X--ray emitting matter is not compatible with
CDM.

We have also used the
non--linear scaling of the power spectrum proposed by PD. This only increases
the excess fluctuations from CDM (by about 50 per cent for the small beam, the
more stringent limit), hence worsening the mismatch. A faster clustering
evolution does not therefore help reconcile CDM with the excess fluctuations
upper limits.

If either $b_{\rm X}<1$ (i.e., the X--ray sources are {\it less} clustered than
the underlying mass distribution) or $f<1$ (i.e., the sources considered in our
XLF do not produce the whole of the XRB), CDM models would be consistent with
our excess fluctuations upper limits, provided that $f\times b_{\rm
X}\approxlt0.7$.

We have already shown that the X--ray sources more clearly associated with
peaks on the matter distribution (clusters) are not relevant for the excess
fluctuations. However, AGN and NELGs have been shown to be important
contributors to the soft X--ray background (50 to 60 per cent of it has been
resolved into these types of sources), and both populations seem to cluster in
the same comoving scales as `normal' galaxies do (see Boyle \& Mo 1993 for a
study of the clustering of X--ray AGN, Shanks \& Boyle 1994). 
Values of $b_{\rm X}$ between 1 and 8 would be
obtained from the results of Miyaji (1994), with $\Omega$ varying in the above
range. A value of the bias parameter $b_{\rm X}<1$ is therefore very unlikely.

As discussed at the end of Section 4.2, the absolute normalization of the
emissivity of the sources that produce the excess fluctuations $j(z)$ cancels
out. We also commented that, if an important fraction of those sources were
distributed at smaller redshifts than the population considered in the XLF used
here, the calculated excess fluctuations produced would {\it increase}. About
95 per cent of the excess fluctuations from CDM are produced at $z<1$; sources
at higher redshift do not contribute significantly to the excess fluctuations.
From this it follows that a possibility of getting $f\sim0.7$ to reconcile CDM
and XRB fluctuations would be to place the unresolved part of the sources of
the XRB at $z>1$.

%#7

$\Delta I_{clus}/\ixrb$ calculated for a \Ginga\ beam size and a
power law emissivity evolution are given in Table \ref{Table4} for
$p=0,3$ and two different values of the maximum redshift of
integration $z_{\rm max}=1,3$. The minimum redshift was set at 0.05,
changing it to 0.1 did not change the results significantly. For a
comoving evolution $p=0$, the upper limits are exceeded at all values
of $\Omega$, and the maximum fraction contributed by $z<1$ sources is
$f<0.3$. A positive evolution is in principle more plausible, in line
with the soft XLF results quoted above. For $p=3$, about 50 per cent
of the XRB intensity has to come from $z>1$ to reconcile the upper
limits with the CDM excess fluctuations. Alternatively, most (70--90
per cent) of the XRB sources could be nearby, but then the density of
the Universe cannot be low ($\Omega>0.2-0.4$).

%#11

Similar upper limits on $f$ were obtained from studies of the angular
correlation function of the XRB both above and below 2~keV and in
angular scales between 1~arcmin and several degrees (see e.g. Carrera
et al. 1991, So\l tan \& Hasinger 1994). However, the alternative
possibility in those studies of a rapid evolution of the source
clustering would not be consistent with our data, as discussed above.

\subsection{Limits on the density contrast from excess fluctuations}

In this section, we will investigate the density contrast of matter in the
Universe ($\droro$) implied by the upper limits obtained on
the excess fluctuations.
Instead of using a CDM power spectrum, we assume
that the sources of the XRB have a spatial correlation
function $\xi(r)=(r/r_0)^{-1.8}$ with a comoving evolution. By performing
its Fourier transform and substituting in Eq. \ref{dii}, we can translate
the limits on $\deltaiclus/\ixrb$ to limits on the spatial correlation
length $r_0$.

The density contrast $\droro$ in a window $W({\bf r})$ is given by

\begin{equation}
\label{droro2}
(\droro)^2=\int d^3{\bf k} \hat\xi({\bf k}) \hat W^2 ({\bf k})
\end{equation}

\noindent where $\hat W({\bf k})$ is the Fourier transform of the window
function, that we have taken here to be a sphere of radius $R$. For this window
function and a power law correlation function, $\droro$ is also a power law on
$R$: $(\droro)^2\propto (r_0/R)^{1.8}$. We can therefore use the limits on
$r_0$ obtained from Eq. \ref{dii} (with the emissivities discussed in Section
4.2) to constrain $\droro$. The resulting upper
limits on $\droro$ versus $R$ are plotted in Fig. 7, using both our limits on
the excess fluctuations from \ROSAT, and Butcher et al. (1996) results from
\Ginga\ (assuming $p=3$ that gives the more conservative upper limits). We
have used $\Omega=0.1$ in Fig. 7. If instead we use $\Omega=1$,
the limits are 10--20 per cent smaller.

Given the size of the different beams used, this analysis is going to be
sensitive to different sampling radii. We have estimated the relevant
ranges by using a typical angular distance for each beam size involved 
($\sim3'$ for our \ROSAT\ small beam, $\sim12'$ for our large beam, and $\sim
1^\circ$ for \Ginga) and calculating the maximum and minimum separations it
corresponds for the redshift range considered ($z\sim0.05-3$ for \ROSAT, and
$z\sim 0.1-3$ for \Ginga). As we can see in Fig. 7, the smaller angular scale
results are sensitive to spatial distances of the order of 1~Mpc, while the
larger ones are sensitive to a few tens of Mpc.

\begin{figure}
  \vbox to 0cm{\vfil}
\epsfverbosetrue
\epsfysize= 180pt
\epsffile{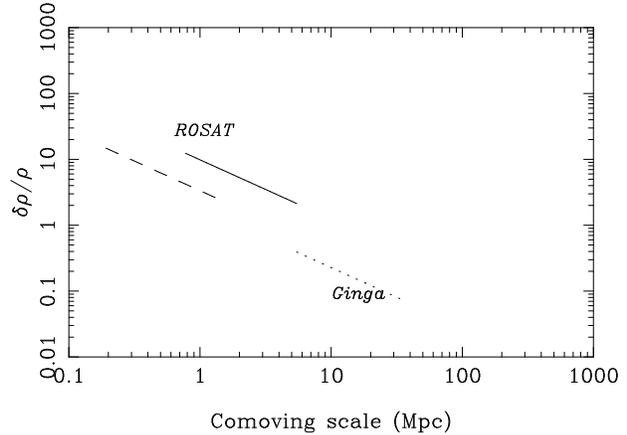}
\caption{$\delta\rho/\rho$ limits at different scales (see text):
the solid line is the \ROSAT\ upper limit from the large beam, the dashed line
is the small beam upper limit, the dotted line is the \Ginga\ upper limit.}
  \label{Fig7}
\end{figure}

At the larger scales sampled here the Universe is quite homogeneous
($\droro<1$), while below a few Mpc there is space for strong density
fluctuations ($\droro>1$), that would reveal a highly non--linear growth of
structure.

{}
{}
{}
{}

\section{SUMMARY}

Our fluctuation analysis of 80 \ROSAT\ fields has allowed us to constrain the
excess fluctuations on $\sim$10~arcmin angular scales to be  $\deltaiclus <
0.004 \, \ctsbeam$ and on $\sim$2~arcmin  $\deltaioclus < 0.0002 \, \ctsbeam$
(or $\deltaiclus/\ixrb<$17 per cent and $\deltaioclus/\ixrb<$7 per cent), both
with 2 sigma confidence levels. 

The source counts found in medium and deep surveys in empty fields
reproduce well the fluctuations of the XRB around bright targets (most
of which are nearby galaxies of different types).  Since there is no
need for any excess fluctuations, we conclude that faint X--ray
sources are not associated to local astronomical objects.

A contribution from extended objects with low surface brightness (like groups
or clusters of galaxies) is not required to fit the observed distribution of
intensities. The surface density of these objects is shown to be $<$3
to 6 times the observed value, limiting the fraction of low surface brightness
sources missed by present surveys.

The upper limits on $\deltaiclus/\ixrb$ obtained here (and others from
the literature) have been compared with CDM theoretical models to
extract constraints on the density parameter of the Universe $\Omega$
and the bias parameter of X--ray emitting sources with respect to the
underlying matter distribution $b_{\rm X}$. Unless $b_{\rm X}\sim0.7$
(which is unlikely), the only possibility for reconciling our results
with CDM would be that the remaining unresolved sources of the soft
XRB (contributing 30 per cent of it) are at $z>1$, and have suffered a
cosmological evolution different from the other known sources of the
soft XRB (AGN and NELGs). %#7 
Similarly, sources that produce about 50
per cent of the 2--10~keV XRB have to be at $z>1$; this fraction could
be larger if $\Omega>0.4$.

In a different approach, a power--law shape is assumed instead for
the spatial correlation function of the XRB sources, constraining the
density contrast to be $<1$ on scales of tens of Mpc and $<10-100$ around
one Mpc. This indicates that the Universe is very homogeneous at
larger scales, but inhomogeneities might be present and common at smaller
scales, as observed in surveys of the nearby Universe.

%%%%
\medskip
\noindent {\bf ACKNOWLEDGEMENTS}
\medskip

\noindent FJC and XB thank the DGICYT for financial support, under project
PB92-00501. ACF thanks the Royal Society for support. FJC thanks E.
Mart\'\i nez--Gonz\'alez for useful discussions. We thank G. Hasinger
(the referee) and T. Miyaji for helpful comments and suggestions that
improved the manuscript.  We would like to thank the RIXOS consortium
for letting us use the positions of their fields. This research has
made use of data obtained from the Leicester Database and Archive
Service at the Department of Physics and Astronomy, Leicester
University, UK.

{}
{}
{}

\bsp

\begin{thebibliography}{99}

\bibitem{b01} Almaini O., Shanks T., Boyle B.J., Griffiths R.E., Roche N.,
Stewart G.C., Georgantopoulos I., 1996, MNRAS, 282, 295

\bibitem{b02} Barber C.R. and Warwick R.S., 1994. MNRAS, 267, 270

\bibitem{b03} Barcons X., 1992, ApJ, 396, 460

\bibitem{b04} Barcons X., Branduardi-Raymont G., Warwick R.S., Fabian
A.C., Mason K.O., McHardy I.M., Rowan-Robinson M., 1994, MNRAS, 268,
833

\bibitem{b05} Barcons X., Fabian A.C., 1988, MNRAS, 230, 189

\bibitem{b06} Bardeen J.M., Bond J.R., Kaiser N. Szalay A.S., 1986,
ApJ, 304, 15

\bibitem{b07} Bevington P.R., 1969, Data Reduction and Error Analysis for
the Physical Sciences, McGraw--Hill, New York

\bibitem{b08} Boyle B.J., Mo H.J., 1993, MNRAS, 260, 925

\bibitem{b09} Boyle B.J., Shanks T., Georgantopoulos I., Stewart G.C.,
Griffiths R.E., 1994, MNRAS, 271, 639

\bibitem{b10} Boyle B.J., McMahon R.G., Wilkes B.J., Elvis M., 1995,
MNRAS, 272, 462

\bibitem{b11} Branduardi-Raymont G. et al., 1994, MNRAS, 270, 947

\bibitem{b12} Butcher J.A., Stewart G.C., Warwick R.S., Fabian A.C.,
Carrera, F.J., Barcons X., Hayashida K., Inoue H., Kii T., 1996,
MNRAS, submitted

\bibitem{b13} Carballo R., Warwick R.S., Barcons X., Gonz\'alez-Serrano
J.I., Barber C.R., Mart\'\i nez-Gonz\'alez E., P\'erez-Fournon I.,
Burgos J., 1995, MNRAS, 277,1312

\bibitem{b14} Carrera F.J., Barcons X., Butcher J., Fabian A.C.,
Stewart G.C., et al., 1991, MNRAS, 249, 698

%
%\bibitem{b01} Carrera F.J., Barcons X., Butcher J., Fabian A.C., Lahav O.,
%Stewart G.C., Warwick R.S., 1995, MNRAS, 275, 22
%

\bibitem{b15} Ciliegi P., Elvis M., Wilkes B.J., Boyle B.J., McMahon R.G.,
1996, MNRAS, submitted

%
%\bibitem{b16} Cole S., Kaiser N., 1988, MNRAS, 233, 637
%

\bibitem{b17} Efstathiou G., Bond J.R., White S.D.M., 1992, MNRAS, 258, 1P
(EBW)

%
%\bibitem{b18} Fabian A.C., Barcons X., 1992, ARA\&A, 34, 429
%

\bibitem{b18} Gendreau K.C. et al., 1995, PASJ, 47, L5

\bibitem{b19} Hasinger G., Turner T.J., George I.M., Boess G., 1992,
NASA/GSFC OGIP Calibration Memo CAL/ROS/92-001

\bibitem{b20} Hasinger G., Burg R., Giacconi R., Hartner G., Schmidt
M., Tr\"umper J., Zamorani G., 1993, A\&A, 275, 1

\bibitem{b21} Inoue H., Kii T., Ogasaka Y., Takahashi T. and Ueda Y., 1996, in:
Zimmermann H.U., Tr\"umper J., Yorke H. (eds.), 
R\"ontgenstrahlung from the Universe, MPE Report, 263, p.323

\bibitem{b22} Jones L.R., et al., 1995, in: Maddox S., Arag\'on--Salamanca A.
(eds.), Wide field spectroscopy and the distant Universe, World Scientific
Press, Singapore, p.346


%
%\bibitem{b01} Lahav O. et al., 1993, Nature, 364, 693
%

\bibitem{b23} Lampton M., Margon B., Bowyer S., 1976, ApJ, 208, 177

%\bibitem{b24} Mason, K.O., et al., 1996, in preparation

\bibitem{b25} McCammon D., Sanders W.T., 1990, ARA\&A, 28, 657

%\bibitem{b26} McHardy I., et al., 1996, in preparation

%\bibitem{b27} Mittaz J.P.D., et al., 1996, in preparation

%
%\bibitem{b01} Miyaji T., Lahav O., Jahoda K., Boldt E., 1994, ApJ, 434, 424
%

\bibitem{b43} Miyaji T., 1994, PhD thesis, University of Maryland

\bibitem{b28} Mulchaey J.S., Davis D.S., Mushotzky R., Burstein D.,
1996, ApJ, 456, 80

\bibitem{b29} Olive K.A., Steigman G., 1995, ApJS, 97, 49


\bibitem{b30} Page M.J., Carrera F.J., Hasinger G., Mason K.O., McMahon
R., Mittaz J.P.D., Barcons X., Carballo R., Gonz\'alez--Serrano I.,
P\'erez--Fournon I., 1996a, MNRAS, 281, 579

\bibitem{b31} Page M.J., Mason K.O., McHardy I.M., Jones L.R., Carrera F.J.,
1996, MNRAS, submitted

\bibitem{b32} Peebles P.J.E., 1980, The Large--Scale Structure
of the Universe, Princeton Univ. Press, Princeton, NJ

\bibitem{b33} Peacock J.A., Dodds S.J., 1994, MNRAS, 267, 1020 (PD)

\bibitem{b34} Piccinotti G., Mushotzky R.F., Boldt E.A., Holt S.S.,
Marshall F.E., Serlemitsos P.J., Shafer R.A., 1982, ApJ, 253, 485

\bibitem{b35} Plucinsky P.P., Snowden S.L., Briel U.G., Hasinger G.,
Pfeffermann E., 1993, ApJ, 418, 519

%
%\bibitem{b01} Rebolo R., 1995, in: Mart\'\i nez--Gonz\'alez E., Sanz
%J.L., (eds.), The Universe at High--z; Large--Scale Structure and the
%CMB, Springer Verlag, p. 98
%

\bibitem{b36} Rees M.J., 1980, in Abell G.O., Peebles P.J.E., eds., Proc.
IAU Symp. 92, Objects at High Redshift. Reidel, Dordrecht, Holland,
p. 209


\bibitem{b37} Romero-Colmenero E., Branduardi-Raymont G., Carrera F.J.,
Jones L.R., Mason K.O., McHardy I.M., Mittaz J.P.D., 1996, MNRAS, 282, 94

\bibitem{b38} Rosati P., Della Ceca R., Burg R., Norman C., Giacconi
R., 1995, ApJ, 445 L11

\bibitem{b39} Scheuer P.A.G., 1974, MNRAS, 166, 329

\bibitem{b40} Shanks, T., Boyle, B.J., 1994, MNRAS, 271, 753

\bibitem{b41} Snowden S.L., Freyberg, M.J., 1993, ApJ, 404, 403

\bibitem{b44} So\l tan A.M., Hasinger G., 1994, A\&A, 288, 77

\bibitem{b45} So\l tan A.M., Hasinger G., Egger R., Snowden S., Tr\"umper J.,
1996, A\&A, 305, 17

\bibitem{b42} Vikhlinin A., Forman W., Jones C., Murray S., 1995, ApJ, 451, 564

%\bibitem{b01} 

\end{thebibliography}
\end{document}